\documentclass[8pt,preprint2]{aastex}

\shorttitle{Time Delays in SDSS~J1004+4112}
\shortauthors{Fohlmeister et al.}

\begin{document}

\title{The Rewards of Patience: An 822 Day Time Delay in the Gravitational Lens SDSS~J1004+4112}
 \author{J. Fohlmeister\altaffilmark{1},
   C. S. Kochanek\altaffilmark{2}, E. E. Falco\altaffilmark{3},
   C. W.  Morgan\altaffilmark{2,4},
   \and J. Wambsganss\altaffilmark{1}
}

   \altaffiltext{1}{Astronomisches Rechen-Institut, Zentrum f\"ur
              Astronomie der Universit\"at Heidelberg, M\"onchhofstr. 12-14,
              69120 Heidelberg, Germany} 
   \altaffiltext{2}{Department of Astronomy, Ohio State University, 140 West
              18th Avenue, Columbus, OH 43210}
   \altaffiltext{3}{Smithsonian Astrophysical Observatory, FLWO, P.O. Box
              97, Amado, AZ 85645} 
   \altaffiltext{4}{Department of Physics, United States Naval Academy, 572C Holloway Road,
              Annapolis, MD 21402}

\begin{abstract}
We present 107 new epochs of optical monitoring data for the four brightest
images of the gravitational lens SDSS J1004+4112 observed between October 2006
and June 2007.  Combining this data with the previously obtained light curves,
we determine the time delays between images A, B and C. We confirm our
previous measurement finding that A leads B by $\Delta t_{BA}=40.6\pm1.8$~days,
and find that image C leads image A by $\Delta\tau_{CA}=821.6\pm2.1$ days. The
lower limit on the remaining delay is that image D lags image A by 
$\Delta\tau_{AD}>1250$ days.  Based on the microlensing of images A and B
we estimate that the accretion disk size at a rest wavelength of 2300\AA\ is 
$10^{14.8\pm0.3}$~cm for a disk inclination of $\cos{i}=1/2$, which is consistent with
the microlensing disk size-black hole mass correlation function given our
estimate of the black hole mass from the MgII line width of $\log M_{BH}/M_\odot=8.44\pm0.14$. 
The long delays allow us to fill in the seasonal gaps and assemble a continuous,
densely sampled light curve spanning 5.7 years whose variability implies a
structure function with a logarithmic slope of $\gamma = 0.35\pm0.02$. 
As C is the leading image, sharp features in the C light curve can be 
intensively studied 2.3 years later in the A/B pair, potentially allowing
detailed reverberation mapping studies of a quasar at minimal cost. 
\end{abstract}

   \keywords{cosmology: observations -- 
             gravitational lensing --
                quasars: individual: (SDSS J1004+4112)
               }

\section{Introduction}

The quasar SDSS~J1004+4112 at $z_s=1.734$ is split into five images by an intervening
galaxy cluster at $z_l=0.68$ \cite{inada,inada2,oguri}. With a maximum image separation of $14\farcs62$, it is
a rare example of a quasar gravitationally lensed by a cluster \cite{wambsganss,inada3}. 
One of the most interesting applications of this system is to use the time delays between
the lensed images to study the structure of the cluster.  If we assume the Hubble 
constant is known, then the delays break the primary model degeneracy of lensing studies
(the ``mass sheet degeneracy''), and the delay ratios constrain the structure even if
the Hubble constant is unknown.  After its discovery, several groups modeled the 
expected time delays in SDSS J1004+4112 and their dependence on the mean mass profile of 
the cluster \cite{kawano,oguri,williams}.   When we measured the shortest delay in the
system, between images A and B, we found a longer delay than predicted by the models
(Fohlmeister et al. 2007, hereafter Paper I) where the discrepancy probably arose 
because the models included the cD galaxy and the cluster halo but neglected 
the significant perturbations from the member galaxies.   As we measure the longer
delays, where the cluster potential should be relatively more important than for
the merging A/B image pair, we would not expect cluster substructures to play
as important a role.

We also expect this lens to have a fairly short time scale for microlensing variability
created by stars either in the intracluster medium or in galaxies near the images.
The internal velocities of a cluster are much higher than in a galaxy (700~km/s
versus 200~km/s), and  SDSS~J1004+4112's position on the sky is almost orthogonal 
to the CMB dipole (Kogut et al. 1993), giving the observer a projected motion on the lens
plane of almost 300~km/s.  In Paper I, we detected microlensing of the continuum emission of the 
A/B images in Paper I and there is also evidence for microlensing of the CIV broad line
\cite{richards,lamer,gomez}.  Once we have measured the time delays we can remove
the intrinsic quasar variability and use the microlensing variability to estimate the mean 
stellar mass and stellar surface density, the transverse velocities, and the 
structure of the quasar source \cite{gilmerino,mortonson,poin,morgan}. 

Finally, we note that SDSS~J1004+4112 could be an ideal laboratory for studying correlations
in the intrinsic variability of quasars.  With, image C leading images
A and B by 2.3 years, sharp variations in image C can be used to plan intensive monitoring
of images A and B to measure the response times as a function of wavelength
(e.g. Kaspi et al. 2007), with the additional advantage that the delay between A and B provides 
redundancies that
protect against weather, the Moon and the Sun.  The long delays between the images also
mean that seasonal gaps are completely filled, and we can examine the structure function
of the variability with a densely-sampled, gap-free light curve (modulo corrections for microlensing).
Such data generally do not exist, since most time variability data for quasars (other
than nearby reverberation mapping targets, e.g. Peterson et al. 2004) have very sparse sampling (e.g. Hawkins 2007
on long time scales for a small number of objects or Vanden Berk et al. 2004 on shorter time scales
for many objects).  

In Paper I \cite{fohli} we presented three years of optical monitoring data for the four 
brightest images of SDSS J1004+4112 spanning 1000 days from December 2003 to June 2006. 
The fifth quasar image, E, is too faint to be detected in our observations. We measured
the time delay between the A and B image pair to be $\Delta\tau_{BA}=38.4\pm2.0$ days. 
While larger separation lenses tend to have longer time delays, for these two images the 
propagation time difference is small, because they form a close image pair ($3\farcs8$) 
from the source lying close to a fold caustic.  For the more widely separated
C and D images we could only estimate lower limits on the delays of 560 and 800
days relative to image B and A.  In this paper we present the 107 new optical monitoring epochs 
for the 2006/2007 season in \S2.  When combined with our previous data we have light
curves spanning 1250 days that allow us to measure the AC delay in \S3.  In \S4
we use the microlensing variability of the A/B images to measure the size of the
quasar accretion disk, and in \S5 we measure the structure function of the intrinsic
variability.  We discuss the future prospects for exploiting this system in \S6.    

\begin{figure*}[t]
   \centering
   \includegraphics[bb= 30 100 520 700, width=10cm,angle=0,clip]{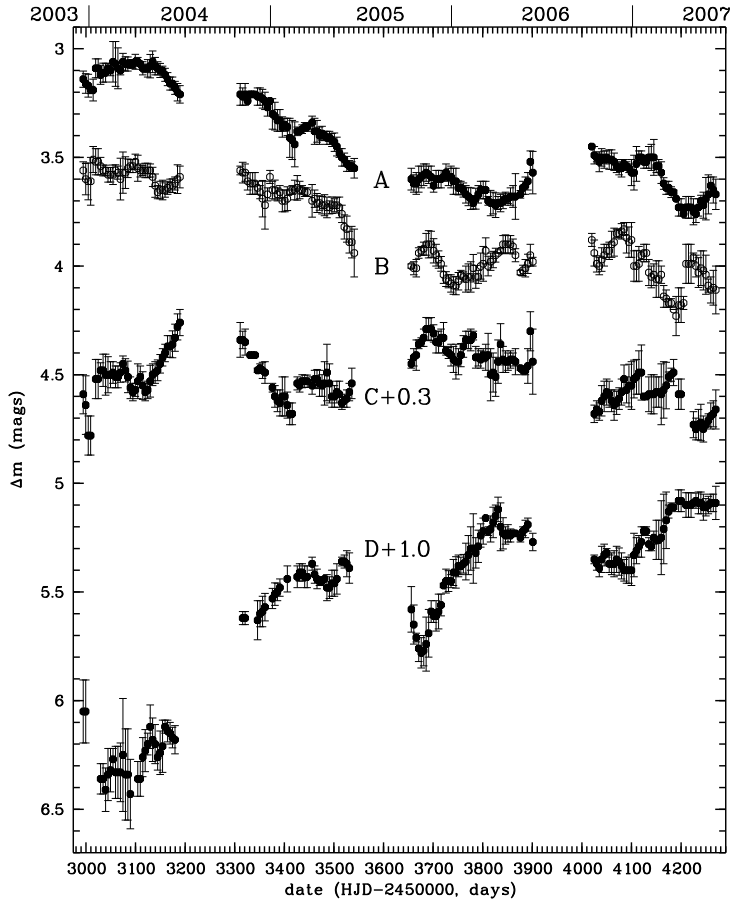}
   \caption{
   Light curves of the A, B, C and D images of the quasar SDSS J1004+4112 from 
   December 2003 to June 2007. Images C and D have been offset by 0.3
   and 1.0 mag, respectively, in order to avoid overlap. We present a running 
   average of one point every 5 days averaged over $\pm7$ days to emphasize
   trends and to avoid confusion by noise.   
   \label{lcurve}} 
\end{figure*}

\section{Data}

We monitored SDSS~J1004+4112 in the r-band during the 2006-2007 season using the Fred 
Lawrence Whipple Observatory (FLWO) 1.2m telescope on Mount Hopkins and the 
MDM 2.4m Hiltner Telescope on Kitt Peak.  The FLWO observations were obtained with 
Keplercam (0\farcs672 pixels) and the MDM observations with  RETROCAM \cite{morgan1}
(0\farcs259 pixels).  The data reduction was carried out as described in Paper I.
We continued to use the same five stars to set the PSF model and the flux 
scale of each epoch and verified that these flux standards continue to show
no variability. Table~1 presents the photometry for the four images in the
2006-2007 season. 

In Figure \ref{lcurve} we present the resulting light curves for images A to D for 
the period from December 2003 to June 2007.  The average sampling rate during the 
2006/2007 season is once every third day.  
The FLWO data are noisy, so for Figure \ref{lcurve} we show a running average 
of the data  (one point every five days averaged over $\pm$ 7 days) to emphasize
the long term trends.  Images C and D are offset by 0.3 and 1.0 mag, respectively, 
so that they do not overlap with image B in the third and fourth season.  During
this season, images A and B faded by approximately 0.4 mag with a prominent feature
near the middle of the season, image C was relatively constant and image D 
brightened by about 0.4 mag.  For the full four seasons, A and B have faded by
approximately 1~mag, C has remained relatively constant and D has brightened
by about 1.5~mag. 

\section{The Time Delay}

For the determination of the time delay, we use the methods described
in Paper I. Our first step with the new data was to remeasure the A/B
delay. The fourth season shows a nice feature with maxima in images
A and B near days 4120 and 4080 respectively, followed by a roughly 100 day 
decline to minima at 4220 (A) and 4180 (B) days. 
With the dispersion method \cite{pelt1,pelt2}
we measure the delay between A and B to be $\Delta t_{BA}=40.1\pm3.5$~days.
For the Kochanek et al. (2006) polynomial method we used 
polynomial orders of $N_{src}=20$, $40$, $60$ and $80$ for the
source and $N_\mu=1$, $2$, $3$ and $4$ for the microlensing 
variability and derived the final estimate using the Bayesian
weighting of these cases described in Poindexter et al. (2007).
We found delays of $40.6\pm1.8$, $40.1\pm1.8$ and $39.8\pm1.8$
(68\% confidence regions) depending on whether we weighted the changes 
in the number of parameters using the Bayesian information criterion (which strongly
penalizes extra parameters), the Akaike information criterion (which
weakly penalizes extra parameters) or no penalty for extra parameters.
These are consistent with our result from Paper I of $\Delta\tau_{BA}=38.4\pm2.0$ 
days, but are somewhat more conservative in their treatment of the
parameterization and the role of microlensing. 

In Paper I we derived a lower limit on the BC delay of $\Delta\tau_{CB}>560$ days
and suggested, based on some similarities between the third season for A/B with
the first season for C, that a delay of order 700 days was plausible but 
statistically too weak to claim as a measurement.  We now see that the
feature in the second season for image C strongly matches the feature
we observe in the new season for A and B.  Using the dispersion 
spectra method (Pelt et al. 1994, 1996), we find $\Delta\tau_{CA}=822\pm7$ days
and  $\Delta\tau_{CB}=780\pm6$ days where the CA delay is slightly less
accurate because the CA overlap is slightly less than the CB overlap due
to the alignment of the light curves relative to the seasonal gaps.  The
three delays are mutually consistent since 
$\Delta\tau_{CB}=\Delta\tau_{CA}-\Delta\tau_{BA}=782\pm7$ days.  For
the polynomial method analysis we simultaneously fit A, B and C
holding the A/B delay fixed to 40.6~days to find CA delays of 
$821.6\pm2.1$, $823.0\pm2.1$ and $820.2\pm2.1$~days for the
three weighting methods, respectively.  
Image D should lag the other three images, and we see no feature in the
light curve of image D that can be matched to the first season of images
A/B.   The lower limit on the time delay between images A and D is
now $\Delta\tau_{DA}>1250$ days (3.4 years).

We modeled the lens using the same approach as in Paper I, where we include
the central cD galaxy, and NFW halo for the cluster dark matter and 12
pseudo-Jaffe models corresponding to cluster
galaxies (we added an extra component at $(x,y)=(31\farcs0,4\farcs0)$ relative
to quasar image A in an effort to reduce the overall shear).  The 
fits were carried out using {\it lensmodel} (Keeton 2001) and while
adequate they are not satisfactory -- it is very difficult to find
solutions with no additional quasar images created by
the galaxies, and checking for the extra images makes the procedure
extraordinarily slow.  At present we lack the ability to model this
system in detail (including uncertainties) at the precision of the
constraints, while simplified models that ignore the galaxies are
incapable of fitting the data at all.  The model predicts an AD delay 
of order 2000~days (5.5 years), which is consistent with our current 
lower bound.

\section{Microlensing and the Size of the Quasar Accretion Disk}

\begin{figure}[t]
   \centering
   \includegraphics[width=7.5cm,angle=0,clip]{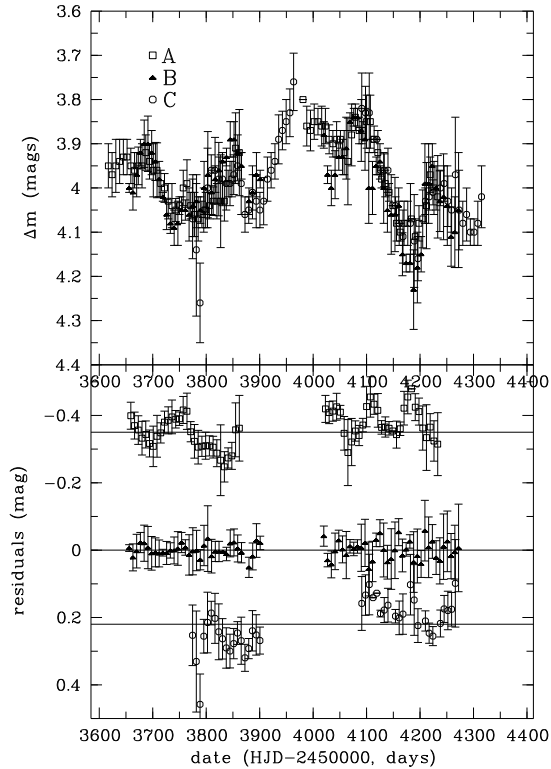}

   \caption{
    The image A, B and C light curves in their overlap region
    after shifting by the time delays. The data are binned in one week
    intervals. The lower box shows the residual magnitudes
    shifted by the offset between the images, revealing microlensing
    variability of order 0.15 mag.
    The light curve of image B was chosen to have constant flux because it
    has the most overlap with the over two.  
   \label{delay}} 
\end{figure}

The residuals of the A and B light curves (see Fig. \ref{delay}) clearly 
indicate that microlensing is present.   After correcting for the time delay,
the mean magnitude differences 
between A and B for the four seasons are $0.460 \pm 0.005$, $0.283\pm0.007$,
$0.339\pm0.005$ and $0.381\pm0.007$~mag.  For the two seasons overlapping
with C we find mean magnitude differences, seasonal gradients and second derivatives of 
$0.590\pm0.010$~mag, $-0.04\pm0.02$~mag/year and $0.29\pm0.09$~mag/year$^2$
for C relative to A and $0.368\pm0.005$~mag, $0.05\pm0.01$~mag/year and
$0.18\pm0.04$~mag/year$^2$ for B relative to A.  Fig.~\ref{delay} shows
the superposition of the phased A, B and C light curves and the differences
between them that are the signature of microlensing. 

We modeled the microlensing for images A/B using the Bayesian Monte Carlo
method of Kochanek (2004).  We used the microlensing parameters of our
(adequate) lens model, with convergence $\kappa$ and shear $\gamma$ 
values of $\kappa=0.48$ and $\gamma=0.57$ for A and 
$\kappa=0.47$ and $\gamma=0.39$ for B.  We allowed the surface
density in stars $\kappa_*$ to vary from 10\% to 100\% of $\kappa$
increments of 10\%.  We used a microlens mass function with 
$dn/dM \propto M^{-1.3}$ with a dynamic range in mass of a factor
of 50 that approximates the Galactic disk mass function of Gould (2000).  
We generated $4096 \times 4096$ pixel magnification patterns with 
an outer scale of $20\langle R_E\rangle$ where $\langle R_E\rangle$
is the Einstein radius at the mean stellar mass $\langle M\rangle$.
We modeled the disk as a face-on, thin disk \cite{shakura}
 neglecting the central temperature depression and
relativistic effects.  We measure the disk size $R_\lambda$ as the point
where the disk temperature matches the rest-frame energy of
our monitoring band, $k T_\lambda = hc/\lambda$, where 
$\lambda\simeq 2300$\AA\ for the r-band at the source redshift
(see Morgan et al. 2007).
The half-light radius $R_{1/2}=2.44R_\lambda$ should be used to
compare to any other disk model, since Mortonson et al. (2005)
have shown that the half-light radius depends little on the
surface brightness profile of the model.  We made four realizations
of each of the 10 microlensing models and drew $2\times 10^5$ trial
light curves for each of the 40 cases so that we would have a
reasonable statistical sampling of light curves that fit the 
data well.  We found that  
\begin{equation}
    R_{2300\AA} = 10^{14.8\pm 0.3} {\hbox{cm} \over h_{70} \sqrt{cos i}}
\end{equation}
for a disk inclination angle $i$,
whether or not we use a prior on the mean microlens mass of
 $0.1M_\odot < \langle M\rangle < M_\odot$.  

\begin{figure}[t]
   \centering
   \includegraphics[width=7.5cm,angle=0,clip]{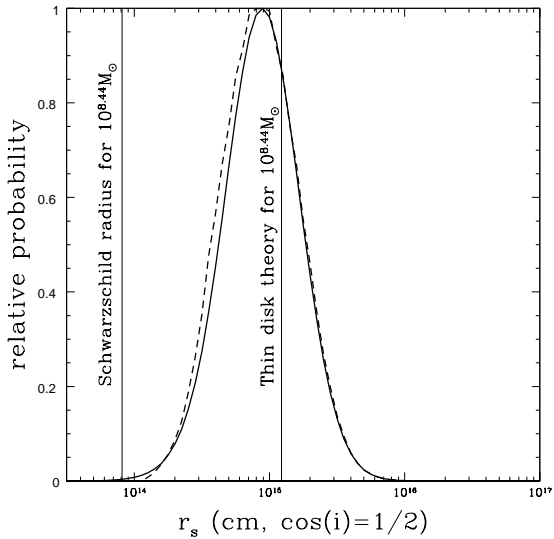}

   \caption{ Probability distribution for the accretion disk size at 2300\AA\
      assuming the mean disk inclination ($\cos(i)=1/2$).  The
      solid (dashed) curves show the distribution without (with) the prior on the
      average microlens mass.  The vertical lines show the Schwarzschild radius
      and the expected size for thin disk theory assuming the black hole mass
      estimated from the MgII emission line width.  The expectation from thin
      disk theory assumes the disk is radiating at Eddington ($L/L_E=1$) with
      efficiency $\eta=0.1$ where $L=\eta \dot{M} c^2$.
   \label{disk}} 
\end{figure}

From the MgII emission line width/black hole mass calibration of 
Kollmeier et al.  (2006), the spectrum of image C from Richards 
et al. (2004), and a magnification-corrected HST $I$-band magnitude 
of $20.9\pm0.4$, we estimate a black hole mass of 
$\log M_{BH}/M_\odot =8.4\pm0.2$.  Fig.~\ref{disk} compares the 
disk size estimate to the characteristic scales of such a black hole.

\section{The Structure Function}

The quasar structure function can be used as a tool to characterize  quasar
variability independent of short-timescale monitoring gaps and to compare
with theoretical models of quasar variability (e.g. Kawaguchi et al. 1998).  
The structure function 
   \begin{equation}
     S(\tau) =\sqrt{{1\over N({\tau})} \sum\limits_{i<j} [m(t_{j})-m(t_{i})]^{2}}
   \end{equation}
is the variance in the magnitude as a function of the time $\tau=t_j-t_i$ between 
measurements where $m(t_{j})$ and $N({\tau})$ is the number of epochs at that 
time lag.  For SDSS J1004+4112 we can determine the structure function over a 
moderate time range and with a dense sampling rate and no seasonal gaps if we
use the time-delay corrected quasar light curves for images B and C.  These 
cover a time-baseline of 2065 days (5.7 years) in the observers frame, corresponding to a maximum
rest-frame time lag at $z_s=1.734$ of 755 days. For the very different behavior of the image
D light curve, which could not yet be time-delay connected to the other images, we
compute the structure function independently for rest frame time lags up to
470 days. As in Vanden Berk et al. 2004 we fit the form of the structure function with a
power law. The value for the power law index $\gamma = 0.35\pm0.02$
for the combined image B and C light curves is consistent with that derived
for the SDSS quasar sample. For image D we find a similar slope of
$\gamma = 0.39\pm0.03$, as expected from the light curve. 
Time-delay
connecting the image A, B and C lightcurves by subtracting the estimated
microlensing variability in the overlap region of the lightcurves gives a
restframe record of the intrinisic quasar variability over 500 days. The slope
of the structure function for the source light curve $\gamma_{s} = 0.45\pm0.03$ is steeper than for the
observed non-microlensing corrected curves.  

\section{Summary and Conclusions}

We present a fourth season of monitoring data for the four bright images of the
five image gravitational lens system SDSS J1004+4112.  We confirm our previous
estimate for the time delay between the merging A/B pair, finding that B leads
A by $40.6\pm1.8$ days.  We measure the delay for image C for the first time,
finding that it leads image A by $821.6\pm2.1$~days.  We note that this is
nearly twice the longest previously measured delay (the 417 day delay in 
Q0957+561 \cite{schild, kundic}).  We find a lower
bound that D lags A by more than approximately 1250~days.  Our current mass model predicts 
that D lags A by approximately 2000 days, which is consistent with the present limit.
The fractional uncertainties
in the AB delay are still dominated by sampling and microlensing, while the
fractional uncertainties in the AC delay are dominated by cosmic variance due
to density fluctuations along the line of sight rather than our measurement
uncertainties of 0.3\% (e.g. Barkana 1996).  

A detailed model of this system, including the constraints from the multiply
imaged, higher redshift arcs (Sharon et al. 2005), the X-ray measurements
\cite{ota,lamer}
and a detailed understanding of the uncertainties will
be a challenge.   We lack a completely satisfactory model for the system
at present, in the sense that the modeling process is extraordinarily slow 
due to the ability of the gravitational potentials associated with the 
cluster member galaxies to generate additional but undetected images of the quasar,
making it impossible to carry out a reliable model survey.  The record of models
for this system is discouraging.  As we noted in Paper I, all three model
studies (Oguri et al. 2004; Williams \& Saha 2004;  
Kawano \& Oguri 2006) generically predicted shorter AB delays than the observed
40 days, and that this could be plausibly explained by the absence of 
substructure (i.e. galaxies) in the potential models.  The longer AB-C
and AB-D delays should be less sensitive to substructure.  Oguri et al.
(2004) do not include an estimate of the AB-C delays and have A-D 
delays consistent with our present limits.  The range of B-C delays
in Williams \& Saha (2004) is consistent with our measurement of 820
days, but they predict AD delays shorter than our current lower bound
of 1250~days.  Kawano \& Oguri (2006) predict a range for the longer 
delays over a broad range of mass distributions, none of which match
our delays in detail.  However, models with sufficiently long C-B delays
generally have C-D delays long enough to agree with our present limits.

Based on our present mass model we used the microlensing between the A
and B images to make an estimate of the size of the quasar accretion disk
at 2300\AA\ in the quasar rest frame.  If we convert this to the expected
size at 2500\AA\ assuming the $R_\lambda \propto \lambda^{4/3}$ scaling
for a thin disk and assume the mean disk inclination $\cos (i)=1/2$ the scale 
on which the disk temperature matches the photon energy is $R_{2500\AA}=10^{15.0\pm0.3}$~cm.
Comparisons to other disk models should use the half-light radius which is $2.44$ times larger.
Based on the quasar MgII emission line width we estimate that the black hole 
mass is 
$10^{8.4\pm0.2} M_\odot$. For this mass, the microlensing accretion disk size-black
hole mass correlation found by Morgan et al. (2007) predicts that $R_{2500\AA}=10^{15.3}$~cm,
which is in broad agreement with the measurement.  Further observations, the
inclusion of additional images, and monitoring in multiple bands should
improve these measurements and potentially allow us to determine the mean
surface density in stars near the images $\kappa_*$ and their average mass
$\langle M\rangle$.  Similarly, the ability to construct continuous light
curves of the intrinsic variability and to use image C to provide early
warning of sharp flux changes that can then be intensively monitored in
images A and B may make this system a good candidate for applying 
reverberation mapping techniques to a massive, luminous quasar.  At
present, we already see that the system has a structure function 
typical of quasars.

\begin{acknowledgements}
 We thank all the participating observers at both the Harvard-Smithsonian
 Center for Astrophysics and the MDM Observatory for their support of these
 observations.  This work is also based on observations obtained with the MDM
 2.4m Hiltner and 1.3m McGraw-Hill telescopes, which are owned and
 operated by a consortium consisting of Columbia University, Dartmouth
 College, the University of Michigan, the Ohio State University and
 Ohio University. We thank N.F. Bate for valuable comments and encouragements.
 We also acknowledge support by the European Community's Sixth
 Framework Marie Curie Research Training Network Programme, Contract
 No. MRTN-CT-2004-505183 ``ANGLES".
\end{acknowledgements}

%----------------------------------------------------------------

\clearpage
\begin{deluxetable}{cccccccc}
\rotate
\tablewidth{0pt}
\tablecaption{Light Curves for SDSS J1004+4112 \tablenotemark{*}}
\tablehead{
\colhead{HJD}           & \colhead{$\chi^{2}/N_{dof}$}      &
\colhead{Image A}       & \colhead{Image B}  &
\colhead{Image C}       & \colhead{Image D}  &
\colhead{Observatory} & \colhead{Detector}}
\startdata
4019.006 &1.15 &3.451$\pm$0.027 &3.880$\pm$0.027 &4.332$\pm$0.027 &4.323$\pm$0.027 &MDM &RETROCAM\\
4029.001 &1.99 &3.488$\pm$0.027 &3.952$\pm$0.027 &4.382$\pm$0.027 &4.347$\pm$0.027 &MDM &RETROCAM\\
4031.006 &1.46 &3.529$\pm$0.027 &3.985$\pm$0.027 &4.413$\pm$0.027 &4.361$\pm$0.027 &MDM &RETROCAM\\
4035.026 &0.99 &3.477$\pm$0.027 &4.024$\pm$0.027 &4.326$\pm$0.027 &4.395$\pm$0.027 &MDM &RETROCAM\\
4035.980 &2.20 &3.511$\pm$0.027 &3.997$\pm$0.027 &4.344$\pm$0.027 &4.362$\pm$0.027 &MDM &RETROCAM\\
4039.016 &2.02 &3.576$\pm$0.027 &4.061$\pm$0.027 &4.394$\pm$0.027 &4.434$\pm$0.027 &MDM &RETROCAM\\
4043.955 &0.86 &3.514$\pm$0.027 &3.924$\pm$0.027 &4.249$\pm$0.027 &4.297$\pm$0.027 &MDM &RETROCAM\\
4044.949 &0.95 &3.525$\pm$0.027 &3.930$\pm$0.027 &4.256$\pm$0.027 &4.318$\pm$0.027 &MDM &RETROCAM\\
4045.966 &0.77 &3.472$\pm$0.027 &3.985$\pm$0.027 &4.451$\pm$0.040 &4.274$\pm$0.034 &MDM &RETROCAM\\
4046.966 &1.03 &3.512$\pm$0.027 &3.916$\pm$0.027 &4.271$\pm$0.027 &4.285$\pm$0.027 &MDM &RETROCAM\\
4047.004 &0.47 &3.427$\pm$0.103 &3.915$\pm$0.158 &4.415$\pm$0.252 &4.625$\pm$0.298 &FLWO &Keplercam\\
4048.968 &1.34 &3.518$\pm$0.027 &3.924$\pm$0.027 &4.276$\pm$0.027 &4.337$\pm$0.027 &MDM &RETROCAM\\
4050.008 &0.97 &3.523$\pm$0.027 &3.943$\pm$0.027 &4.245$\pm$0.027 &4.322$\pm$0.027 &MDM &RETROCAM\\
4054.008 &0.59 &3.486$\pm$0.027 &3.877$\pm$0.034 &4.258$\pm$0.049 &4.344$\pm$0.052 &FLWO &Keplercam\\
4059.902 &1.41 &3.580$\pm$0.027 &3.998$\pm$0.027 &4.415$\pm$0.027 &4.393$\pm$0.027 &MDM &RETROCAM\\
4059.947 &1.46 &3.490$\pm$0.027 &3.936$\pm$0.027 &4.291$\pm$0.027 &4.419$\pm$0.027 &FLWO &Keplercam\\
4060.884 &2.58 &3.477$\pm$0.027 &3.908$\pm$0.027 &4.282$\pm$0.027 &4.412$\pm$0.027 &FLWO &Keplercam\\
4065.001 &0.92 &3.498$\pm$0.027 &3.875$\pm$0.028 &4.339$\pm$0.042 &4.330$\pm$0.041 &FLWO &Keplercam\\
4066.006 &0.80 &3.507$\pm$0.048 &3.824$\pm$0.063 &4.313$\pm$0.100 &4.301$\pm$0.098 &FLWO &Keplercam\\
4067.965 &1.40 &(3.655$\pm$0.027) &(4.007$\pm$0.027) &4.487$\pm$0.027 &4.440$\pm$0.027 &MDM &RETROCAM\\
4070.954 &1.16 &3.522$\pm$0.027 &3.835$\pm$0.027 &4.330$\pm$0.041 &4.307$\pm$0.040 &FLWO &Keplercam\\
4071.884 &0.96 &3.566$\pm$0.027 &3.854$\pm$0.027 &4.292$\pm$0.027 &4.306$\pm$0.027 &MDM &RETROCAM\\
4072.014 &1.48 &3.525$\pm$0.027 &3.858$\pm$0.027 &4.323$\pm$0.027 &4.319$\pm$0.027 &FLWO &Keplercam\\
4072.890 &1.08 &3.533$\pm$0.046 &3.820$\pm$0.059 &4.317$\pm$0.094 &4.304$\pm$0.092 &FLWO &Keplercam\\
4072.949 &0.81 &3.554$\pm$0.027 &3.867$\pm$0.027 &4.298$\pm$0.027 &4.385$\pm$0.027 &MDM &RETROCAM\\
4073.894 &0.61 &3.585$\pm$0.027 &3.865$\pm$0.027 &4.345$\pm$0.035 &4.332$\pm$0.035 &MDM &RETROCAM\\
4074.897 &1.27 &3.573$\pm$0.027 &3.841$\pm$0.027 &4.285$\pm$0.027 &4.301$\pm$0.027 &MDM &RETROCAM\\
4074.996 &1.02 &3.509$\pm$0.037 &3.873$\pm$0.051 &4.350$\pm$0.079 &4.390$\pm$0.081 &FLWO &Keplercam\\
4075.993 &0.86 &3.550$\pm$0.041 &3.797$\pm$0.051 &4.310$\pm$0.081 &4.464$\pm$0.093 &FLWO &Keplercam\\
4076.029 &1.11 &3.604$\pm$0.027 &3.848$\pm$0.027 &4.349$\pm$0.027 &4.313$\pm$0.027 &MDM &RETROCAM\\
4081.022 &1.29 &3.550$\pm$0.033 &3.866$\pm$0.043 &4.178$\pm$0.057 &4.436$\pm$0.072 &FLWO &Keplercam\\
4082.865 &1.23 &3.521$\pm$0.028 &3.802$\pm$0.034 &4.192$\pm$0.049 &4.334$\pm$0.055 &FLWO &Keplercam\\
4084.972 &0.82 &3.506$\pm$0.027 &3.844$\pm$0.027 &4.280$\pm$0.027 &4.390$\pm$0.027 &FLWO &Keplercam\\
4086.877 &0.91 &3.524$\pm$0.027 &3.826$\pm$0.034 &4.242$\pm$0.049 &4.424$\pm$0.057 &FLWO &Keplercam\\
4092.985 &0.62 &3.557$\pm$0.031 &3.944$\pm$0.043 &4.256$\pm$0.057 &4.427$\pm$0.067 &FLWO &Keplercam\\
4093.956 &3.20 &3.527$\pm$0.027 &3.782$\pm$0.033 &4.250$\pm$0.051 &4.330$\pm$0.054 &FLWO &Keplercam\\
4094.999 &0.62 &3.539$\pm$0.027 &3.849$\pm$0.035 &4.242$\pm$0.051 &4.451$\pm$0.061 &FLWO &Keplercam\\
4096.038 &2.08 &(3.692$\pm$0.027) &3.990$\pm$0.027 &4.408$\pm$0.027 &4.454$\pm$0.027 &MDM &RETROCAM\\
4096.050 &1.25 &3.567$\pm$0.027 &3.870$\pm$0.028 &4.214$\pm$0.038 &4.320$\pm$0.041 &FLWO &Keplercam\\
4100.947 &1.00 &3.567$\pm$0.033 &3.892$\pm$0.043 &4.143$\pm$0.055 &4.435$\pm$0.070 &FLWO &Keplercam\\
4102.850 &0.38 &3.673$\pm$0.089 &3.849$\pm$0.104 &4.249$\pm$0.151 &4.204$\pm$0.146 &FLWO &Keplercam\\
4107.900 &0.70 &3.668$\pm$0.061 &4.024$\pm$0.084 &4.232$\pm$0.104 &4.694$\pm$0.153 &FLWO &Keplercam\\
4108.972 &0.86 &3.499$\pm$0.040 &4.007$\pm$0.063 &4.268$\pm$0.082 &4.332$\pm$0.086 &FLWO &Keplercam\\
4109.979 &0.56 &3.547$\pm$0.063 &4.086$\pm$0.101 &4.271$\pm$0.123 &4.494$\pm$0.149 &FLWO &Keplercam\\
4111.021 &1.10 &3.490$\pm$0.034 &3.971$\pm$0.051 &4.244$\pm$0.066 &4.236$\pm$0.066 &FLWO &Keplercam\\
4115.026 &0.58 &3.468$\pm$0.028 &3.923$\pm$0.041 &4.120$\pm$0.049 &4.347$\pm$0.060 &FLWO &Keplercam\\
4117.922 &1.25 &3.519$\pm$0.027 &3.960$\pm$0.031 &4.128$\pm$0.037 &4.259$\pm$0.041 &FLWO &Keplercam\\
4126.888 &1.44 &3.502$\pm$0.027 &3.955$\pm$0.036 &4.308$\pm$0.050 &4.209$\pm$0.046 &FLWO &Keplercam\\
4127.928 &0.82 &3.534$\pm$0.030 &3.904$\pm$0.041 &4.310$\pm$0.059 &4.210$\pm$0.054 &FLWO &Keplercam\\
4128.896 &1.99 &3.513$\pm$0.028 &3.974$\pm$0.041 &4.284$\pm$0.057 &4.248$\pm$0.054 &FLWO &Keplercam\\
4137.930 &1.00 &3.505$\pm$0.042 &4.040$\pm$0.067 &4.357$\pm$0.091 &4.306$\pm$0.086 &FLWO &Keplercam\\
4138.775 &0.71 &3.558$\pm$0.052 &4.106$\pm$0.084 &4.188$\pm$0.095 &4.387$\pm$0.112 &FLWO &Keplercam\\
4139.857 &1.68 &3.411$\pm$0.049 &3.991$\pm$0.082 &4.143$\pm$0.098 &4.340$\pm$0.115 &FLWO &Keplercam\\
4140.820 &1.39 &3.507$\pm$0.027 &4.002$\pm$0.027 &4.336$\pm$0.036 &4.273$\pm$0.035 &FLWO &Keplercam\\
4150.876 &1.36 &3.543$\pm$0.028 &4.026$\pm$0.041 &4.348$\pm$0.055 &4.229$\pm$0.050 &FLWO &Keplercam\\
4152.757 &0.67 &3.533$\pm$0.038 &4.167$\pm$0.067 &4.355$\pm$0.078 &4.420$\pm$0.084 &FLWO &Keplercam\\
4153.758 &0.59 &3.580$\pm$0.033 &4.007$\pm$0.047 &4.233$\pm$0.060 &4.183$\pm$0.057 &FLWO &Keplercam\\
4155.833 &0.58 &3.499$\pm$0.038 &4.094$\pm$0.064 &4.187$\pm$0.072 &4.252$\pm$0.076 &FLWO &Keplercam\\
4156.826 &0.89 &3.565$\pm$0.035 &4.009$\pm$0.051 &4.271$\pm$0.066 &4.211$\pm$0.062 &FLWO &Keplercam\\
4165.922 &0.63 &3.625$\pm$0.027 &(4.455$\pm$0.027) &4.535$\pm$0.027 &4.534$\pm$0.027 &FLWO &Keplercam\\
4166.764 &1.84 &3.604$\pm$0.040 &4.065$\pm$0.059 &4.221$\pm$0.070 &4.071$\pm$0.062 &FLWO &Keplercam\\
4168.830 &0.50 &3.659$\pm$0.041 &4.135$\pm$0.061 &4.219$\pm$0.068 &4.176$\pm$0.066 &FLWO &Keplercam\\
4169.811 &1.40 &3.632$\pm$0.027 &4.151$\pm$0.040 &4.231$\pm$0.044 &4.180$\pm$0.042 &FLWO &Keplercam\\
4170.776 &0.52 &3.603$\pm$0.027 &4.156$\pm$0.027 &4.181$\pm$0.028 &4.143$\pm$0.027 &FLWO &Keplercam\\
4171.800 &1.45 &3.612$\pm$0.027 &4.193$\pm$0.027 &4.223$\pm$0.027 &4.139$\pm$0.027 &FLWO &Keplercam\\
4172.714 &0.95 &3.639$\pm$0.027 &4.173$\pm$0.039 &4.212$\pm$0.041 &4.117$\pm$0.038 &FLWO &Keplercam\\
4173.754 &2.61 &3.636$\pm$0.027 &4.147$\pm$0.027 &4.233$\pm$0.027 &4.120$\pm$0.027 &FLWO &Keplercam\\
4174.778 &0.99 &3.665$\pm$0.027 &4.155$\pm$0.036 &4.208$\pm$0.039 &4.099$\pm$0.035 &FLWO &Keplercam\\
4176.850 &0.89 &3.692$\pm$0.027 &4.157$\pm$0.038 &4.207$\pm$0.041 &4.137$\pm$0.039 &FLWO &Keplercam\\
4177.700 &1.10 &3.635$\pm$0.027 &4.166$\pm$0.034 &4.176$\pm$0.035 &4.082$\pm$0.033 &FLWO &Keplercam\\
4179.665 &0.98 &3.651$\pm$0.027 &4.194$\pm$0.036 &4.217$\pm$0.038 &4.107$\pm$0.035 &FLWO &Keplercam\\
4180.687 &0.91 &3.664$\pm$0.027 &4.200$\pm$0.038 &4.162$\pm$0.038 &4.120$\pm$0.037 &FLWO &Keplercam\\
4194.799 &0.68 &3.688$\pm$0.054 &4.229$\pm$0.087 &4.266$\pm$0.093 &4.046$\pm$0.076 &FLWO &Keplercam\\
4197.724 &1.03 &3.734$\pm$0.027 &4.138$\pm$0.035 &4.230$\pm$0.039 &4.059$\pm$0.034 &FLWO &Keplercam\\
4201.761 &0.76 &3.763$\pm$0.027 &4.167$\pm$0.027 &4.361$\pm$0.027 &4.139$\pm$0.027 &FLWO &Keplercam\\
4213.802 &0.63 &3.739$\pm$0.057 &4.046$\pm$0.075 &4.371$\pm$0.105 &4.115$\pm$0.083 &FLWO &Keplercam\\
4214.699 &0.93 &3.675$\pm$0.064 &3.892$\pm$0.078 &4.425$\pm$0.131 &4.066$\pm$0.095 &FLWO &Keplercam\\
4215.700 &0.80 &3.782$\pm$0.060 &4.041$\pm$0.076 &4.507$\pm$0.118 &4.113$\pm$0.083 &FLWO &Keplercam\\
4227.656 &0.50 &3.724$\pm$0.043 &3.947$\pm$0.052 &4.471$\pm$0.084 &4.085$\pm$0.061 &FLWO &Keplercam\\
4230.661 &0.85 &3.778$\pm$0.035 &3.976$\pm$0.041 &4.390$\pm$0.060 &4.098$\pm$0.047 &FLWO &Keplercam\\
4232.706 &0.56 &3.800$\pm$0.047 &4.043$\pm$0.058 &4.517$\pm$0.091 &4.061$\pm$0.062 &FLWO &Keplercam\\
4233.737 &0.69 &3.720$\pm$0.039 &4.054$\pm$0.052 &4.436$\pm$0.074 &4.065$\pm$0.054 &FLWO &Keplercam\\
4237.673 &0.54 &3.700$\pm$0.038 &4.016$\pm$0.050 &4.319$\pm$0.067 &4.081$\pm$0.054 &FLWO &Keplercam\\
4238.688 &0.58 &3.713$\pm$0.042 &3.976$\pm$0.052 &4.485$\pm$0.084 &4.114$\pm$0.061 &FLWO &Keplercam\\
4239.745 &0.45 &3.682$\pm$0.050 &4.147$\pm$0.075 &4.380$\pm$0.094 &4.177$\pm$0.080 &FLWO &Keplercam\\
4240.707 &0.75 &3.721$\pm$0.038 &3.985$\pm$0.048 &4.491$\pm$0.076 &4.039$\pm$0.052 &FLWO &Keplercam\\
4245.660 &0.88 &3.715$\pm$0.053 &4.039$\pm$0.070 &4.464$\pm$0.105 &4.073$\pm$0.074 &FLWO &Keplercam\\
4246.714 &0.34 &3.662$\pm$0.076 &3.887$\pm$0.092 &4.249$\pm$0.133 &4.149$\pm$0.121 &FLWO &Keplercam\\
4247.702 &2.05 &3.788$\pm$0.086 &4.055$\pm$0.109 &4.247$\pm$0.133 &4.130$\pm$0.120 &FLWO &Keplercam\\
4248.668 &0.77 &3.759$\pm$0.069 &4.235$\pm$0.104 &4.763$\pm$0.169 &4.233$\pm$0.107 &FLWO &Keplercam\\
4249.684 &1.54 &3.850$\pm$0.081 &4.358$\pm$0.126 &4.377$\pm$0.133 &4.198$\pm$0.113 &FLWO &Keplercam\\
4250.685 &0.87 &3.675$\pm$0.062 &4.109$\pm$0.091 &4.641$\pm$0.147 &4.164$\pm$0.098 &FLWO &Keplercam\\
4252.675 &1.74 &3.531$\pm$0.067 &4.010$\pm$0.103 &4.445$\pm$0.153 &4.085$\pm$0.113 &FLWO &Keplercam\\
4254.660 &0.55 &3.652$\pm$0.035 &4.077$\pm$0.049 &4.426$\pm$0.069 &4.068$\pm$0.051 &FLWO &Keplercam\\
4255.667 &1.34 &3.651$\pm$0.034 &4.081$\pm$0.049 &4.431$\pm$0.068 &4.119$\pm$0.052 &FLWO &Keplercam\\
4258.675 &0.50 &3.622$\pm$0.040 &4.104$\pm$0.060 &4.359$\pm$0.076 &4.144$\pm$0.064 &FLWO &Keplercam\\
4260.697 &0.73 &3.639$\pm$0.036 &4.045$\pm$0.050 &4.395$\pm$0.069 &4.062$\pm$0.052 &FLWO &Keplercam\\
4261.716 &0.51 &3.623$\pm$0.040 &4.114$\pm$0.060 &4.422$\pm$0.082 &4.065$\pm$0.060 &FLWO &Keplercam\\
4263.668 &0.48 &3.651$\pm$0.050 &4.231$\pm$0.083 &4.314$\pm$0.092 &4.117$\pm$0.077 &FLWO &Keplercam\\
4264.654 &0.69 &3.615$\pm$0.046 &4.098$\pm$0.069 &4.467$\pm$0.096 &4.032$\pm$0.067 &FLWO &Keplercam\\
4265.685 &0.48 &3.693$\pm$0.046 &4.141$\pm$0.068 &4.301$\pm$0.081 &4.114$\pm$0.069 &FLWO &Keplercam\\
4266.700 &0.73 &3.531$\pm$0.068 &4.243$\pm$0.129 &4.148$\pm$0.124 &4.253$\pm$0.135 &FLWO &Keplercam\\
4269.687 &1.27 &3.704$\pm$0.083 &4.117$\pm$0.120 &4.426$\pm$0.159 &3.972$\pm$0.111 &FLWO &Keplercam\\
4271.674 &0.44 &3.742$\pm$0.077 &3.960$\pm$0.093 &4.317$\pm$0.133 &4.037$\pm$0.104 &FLWO &Keplercam\\
4276.677 &0.36 &3.720$\pm$0.077 &4.101$\pm$0.107 &4.360$\pm$0.140 &4.122$\pm$0.113 &FLWO &Keplercam\\
4277.649 &0.25 &3.695$\pm$0.167 &3.986$\pm$0.213 &4.195$\pm$0.266 &4.277$\pm$0.282 &FLWO &Keplercam\\
4278.668 &0.32 &3.615$\pm$0.079 &4.099$\pm$0.121 &4.576$\pm$0.190 &4.002$\pm$0.116 &FLWO &Keplercam\\
\tableline
\enddata
\tablecomments{The Heliocentric Julian Days (HJD) column gives the date of the observation
  relative to HJD$=2450000$.  
  The $\chi^2/N_{dof}$ column indicates how well our photometric model fit the imaging
  data.  When $\chi^2>N_{dof}$ we rescale the photometric errors presented in this
  Table by $(\chi^2/N_{dof})^{1/2}$ before carrying out the time delay analysis to
  reduce the weight of images that were fit poorly. 
   The image magnitudes are relative to the comparison stars 
  (see text). The magnitudes enclosed in parentheses are not used in the time 
  delay estimates. } 
\end{deluxetable}					

\clearpage

\end{document}